# Macroscopic-Range Correlations in an Asymmetric Quantum Hall System

S.A.Emelyanov

*A.F.Ioffe Institute, 194021 St. Petersburg, Russia*

A local photo-response of an asymmetric quantum Hall system in tilted magnetic field is demonstrated to be a complicate non-monotonic function of *macroscopic* distances to the sample edges. A local photo-response *in the dark* has been observed caused by illumination of the sample arias remote on a *macroscopic* distances. These macroscopic-range correlatons indicate the presence of a macroscopic ordering of electrons in the system.

At present, the only few works are devoted to detailed study of asymmetric quantum Hall systems (QHS) in tilted magnetic fields. This may be related to the fact that this system is very close to the "classic" QHS. The only distinctions are (i) the magnetic field is slightly tilted and (ii) there is a potential gradient across the well, i.e. the so-called "built-in" electric field. At first sight, these distinct features seem to be so insignificant that it seems to be unlikely to observe any fundamental phenomena unknown in the "classic" QHS. On the other hand, from the symmetry point of view a combined effect of both in-plane magnetic field and "built-in" electric field is known to result in the breaking of time-reversal and space-inversion symmetry in the QHS that provides potentiality for new phenomena.

The first attempt to describe QHS under the conditions of time-reversal and space-inversion symmetry breaking has been done in [1]. In this work, Schrodinger equation was solved for an uncertain (in *XY* plane) QHS possessing an arbitrary transverse potential $V(z)$ in presence of magnetic field tilted in *XZ* plane. Perturbation approach has been used in the limit of small tilting angles. The electrons on the Landau levels were supposed to be spatially separated along *X* axis while along *Y* their wave functions were supposed to be Bloch ones. Although the energy spectrum has not been found in this work, it nevertheless has been shown that the Landau level degeneracy may be lifted by the effect of both built-in electric field and in-plane magnetic field. Positions of the centers of the electrons' cyclotron orbits were found to be in a correlation with their wave vectors in accordance with the following formula known in the "classic" QHS (see, e.g., [2]):

$$x_0 = -k_y r^2, \qquad (1)$$

where $x_0$ is electron's coordinate along *X*, $k_y$ is its wave vector along *Y*, $r$ is the magnetic length. The principle difference between "classic" QHS and asymmetric QHS in tilted magnetic fields is that in former case electrons' directed velocity is exactly zero because of the axial symmetry in the well plane. In later case, if the Landau level degeneracy is truly lifted, then $x_0$ appears to be a function of not only the wave vector but also the electrons' velocity along *Y* which is no more exactly zero. Although the idea of a macroscopic ordering of electrons has not been discussed in [1], *de facto* Eq. (1) is nothing but the rule of the ordering of electrons along *X* in accordance with their velocity along *Y*. Moreover, this ordering should be of a macroscopic character because $x_0$ varies in a quite macroscopic range. Note also that owing to the spatial separation of electrons along *X*, they should behave as elementary spontaneous one-dimensional currents along *Y* axis flowing even at the ground state of the QHS. On the other hand, applicability of the model to any real systems is at least an open question. One of the principle problem is that the system is supposed to be uncertain. Vital importance of this point has become evident from the fact that spontaneous currents along *Y* should be essentially one-dimensional and should thus occur only in the systems uncertain along *Y*. In any certain ones, these currents (if any) should clearly be locked and hence should not be one-dimensional.

Detailed experimental study of asymmetric QHS in tilted magnetic fields was focused primarily on the measurements of terahertz-light-induced fast-response ballistic currents in asymmetric InAs single quantum wells under cyclotron resonance (CR) conditions. Surprisingly, even the first observations by this method appear to be in a qualitative agreement with the model of uncertain asymmetric QHS [3]. A light-induced current along *Y* axis has been observed in *unbiased* structures at $\hbar\omega_c = 13.7$ meV when the electron motion should be restricted in all directions. This fact could be interpreted in terms of a nonzero electrons' velocity on the Landau levels. Further experiments with either two or three pairs of identical spatially separated ohmic contacts have demonstrated that a local light-induced current along *Y* is sensitive to the position of the contacts along *X* axis that could be interpreted in terms of a macroscopic ordering of electrons along *X* [4]. However, something later several works appear in which an alternative explanation has been proposed related to the presence of a small fraction of two-dimensional holes in the system (see, e.g., [5]). These holes were supposed to be responsible for a potential gradient along *X* which could provide a monotonic dependence of local light-induced current vs *X* coordinate.

In this work, we have directly demonstrated that in asymmetric QHS under magnetic fields tilted in *XZ* plane a local light-induced ballistic current along *Y* axis is a complicate non-monotonic function of macroscopic distances to the sample edges along *X* axis. We have also observed the effect of a local photo-response in the dark caused by illumination of the sample arias remote on a macroscopic distance. These macroscopic-range correlations indicate the presence of a macroscopic ordering of electrons in the system.

MBE-grown not-intentionally doped InAs/GaSb single quantum-well structures have been used in the experiments. They consisted of 15 nm of InAs sandwiched between two 10-nm AlSb barriers to avoid hybridization-related effects and capped by a 25-nm GaSb protecting layer. Low-temperature electron density and mobility are $1.4 \bullet 10^{12}$ cm$^{-2}$ and $10^5$ cm$^2$/Vs, respectively. Growth parameters were those which provide the built-in electric field caused primarily by ionised donors on the lower interface of the well so that the field is pointed toward the sample surface. Underestimation of this field yields the value of about $10^4$ V/cm that is consistent with the commonly used values [6]. All samples are supplied by indium ohmic contacts. Geometry of the contacting will be discussed separately.

We study terahertz-light-induced in-plane currents in unbiased structures in presence of quantizing magnetic field tilted on the angle á $= 15°$. As a source of terahertz light we use pulsed NH$_3$ gas laser optically pumped by CO$_2$ laser. The laser wavelength is 90.6 μm ($\hbar\omega = 13.7$ meV) while pulse duration and radiation intensity are 40 ns and 100 W/cm$^2$, respectively. Linearly polarized laser light is normal to the sample surface. In-plane current pulses are monitored by a high-speed storage oscilloscope through the voltage drop on a 50 Ohm load resistor in a short-circuit regime. The experiments are performed at $T = 1.9$ K. To avoid experimental errors related to an uncontrollable variation of built-in field, below we will use only the samples from the same wafer.

We start from the measurements of total light-induced current as a function of magnetic field. Experimental curve is shown in Fig. 1 while geometry of the experiment is shown in the inset. Two branches on the curve are clearly seen. The low-field branch (below 2.5 T) is a non-resonant ohmic current which tends to disappear at higher magnetic fields due to the drop of ohmic conductivity. At high-field branch (above 3 T), the current shows a resonant behaviour reaching the maximum value at about 5 T. Rigorously, this value is higher than that observed in transmission spectra (see, e.g., [7]). However, this fact could be explained by the effect of InAs conduction band non-parabolicity in the high-density electron gas. As for the resonance broadening, one of the potential reasons for its relatively high width is the saturation of CR absorption which should occur at relatively low intensities [8]. Detailed analysis of both position and line-shape of the resonance should be the matter of a separate study. For us it is important only that the resonance is truly CR-related and below we will concentrate primarily on qualitative experiments which could clarify the basic processes in the system studied. Note only that switching of $B$ from $B = B^+$ to $B = B^-$ results in reversing of the current while at $\vec{a} = 0$ the current is exactly zero in accordance with general symmetry relations.

To find a local light-induced ballistic current as a function of macroscopic distance to the sample edges along $X$ under uniform exposition of the whole sample, we have done the experiment sketched in Fig. 2a. We use nine pairs of short ohmic contacts (1 mm in length) lined-up along $X$ axis so that the distance between neighbouring pairs is also 1 mm. To avoid any edging effects, the distances between the contacts and sample edges are not shorter than 0.5 mm. We measure the current through each contact pair at CR point ($B = 5$ T). At this point, the load resistance is at least two order lower than the resistance between nearest contacts. The results for two opposite magnetic fields are shown in Figs. 2b and 2c. It is seen that the local photoresponse is a non-monotonic and non-periodic function of macroscopic distances to the sample edges which changes the sign at least four times on the sample length. Such a behaviour clearly can not be explained by the presence of holes or by the presence of any gradients along $X$ axis. We thus see no other way as to suppose the presence of a macroscopic ordering of electrons along $X$ axis. Taking also into account that light-induced ballistic currents are known to be determined by the electrons' directed velocity at both initial and final states of the optical transitions, one can suppose the ordering to be related to a correspondence between electrons' velocity along $Y$ and their position along $X$.

In general, the presence of a macroscopic-range ordering of electrons in an asymmetric QHS seems to be so extraordinary that requires more arguments. To this aim, we carry out the experiment sketched in Fig. 3a. Once more, we measure the light-induced current at CR point when the whole sample is exposed to light but here we use only one contact pair centred along $X$ axis. The idea is that we will make several mechanical splits of the sample along $Y$ axis so that the macroscopic distances from the contact pair to the sample edges along $X$ will change stepwise and we will measure the current after each split. The results are shown in Fig. 3b. Dramatic modifications of the current after each split are clearly seen that clearly confirms the idea of a macroscopic ordering of electrons.

The question that may arise in connection with the last experiment is whether the electrons are also sensitive to some other local perturbations remote on a macroscopic distance? To answer this question we have done the following experiment. We take relatively short sample (9 mm in length) supplied by four contact pairs (Fig. 4a). The length of the contacts as well as the distance between them are the same as in the previous experiments. As a first step, we measure the current through each pair under the exposition of the whole sample at $B = B^+$, i.e. just what we have done in the experiment sketched in Fig. 2a. The results are shown in Fig. 4b which are consistent with those shown in Fig. 2b. After that we make the following operation. We cover about two thirds of the sample by a non-transparent metallic mask as it is shown in Fig. 4a and then we repeat the experiment once more. The results exceed even bravest expectations (Fig. 4c). It is clearly seen that despite of the fact that the contact pair No. 4 is exposed to light of the same intensity, the current through it reduces drastically. On the other hand, the current is clearly observed through those pairs which are remote on a macroscopic distance from the exposed aria. Moreover, the current through the *most distant* pair is *higher* than the current through any other ones. Roughly, the system behaves as if we reduce the intensity of exposition of the *whole* sample but not the aria exposed to light. Of course, to make any quantitative conclusions, further detailed experiments are required. However, we anyway can conclude that local light-induced currents in the system are not the exclusive result of exposition of a given local aria but consist of a number of contributions including those from the arias remote on a macroscopic distance.

It seems to be undoubted that various macroscopic-range correlations indicate the presence of a macroscopic ordering in the system. However, the natural question is what about the mechanism of these correlations? The model of uncertain QHS can not provide comprehensive answer on this question. Nevertheless, it contains a "rational kernel" related to the idea of spatially-ordered spontaneous currents in the system. Indeed, according to our experiments the electrons should be spatially separated along $X$ axis and should possess a nonzero velocity along $Y$ correlated with their $X$ coordinate. Otherwise, the local light-induced current should be exactly zero or at least should not be a function of $X$ coordinate. This seems to be quite enough to suppose the presence of spatially-ordered spontaneous currents along $Y$. Moreover, to provide a detectable macroscopic-range current along $Y$, these spontaneous currents should envelope macroscopic arias of the sample. In fact, we thus suppose the system to be in a broken-symmetry phase characterized by the presence of spontaneous currents at the ground state. Such phases are known as the so-called flux phases (see, e.g., [9, 10]). However, the principle distinct feature of the flux phase in the QHS with respect to any known ones is that in former case the characteristic scale of spontaneous currents should be quite macroscopic while in later case this scale is known to be as small as few angstroms.

To be fair, it should be noted that presented results seem to be rather paradoxical than natural. We could outline the following global problems those appear in the context of these results. The first problem is related to the mechanism responsible for the macroscopic-range correlations we observe. Although a key role of a macroscopic ordering in the mechanism seems to be evident, in itself it is still unclear. The second problem is related to the macroscopic-range spontaneous currents in the system. Although the presence of these currents as well as their spatial ordering directly follow from the experiment, the trajectories of these currents as well as the rule of their spatial ordering are also unclear. As a result, the mechanism of the light-induced ballistic currents is not transparent yet. The third problem is related to the absence of any calculations of the energy spectrum of the system. Moreover, we would like to attract special attention to the problem which seems to be important for any further theoretical studies. This is the problem of boundary conditions. As a rule, selection of these conditions is associated with the basic principle according to which the properties of a macroscopic system can not be determined by the boundary conditions. This principle is commonly used in the "classic" QHS. However, as it directly follows from our experiments, utilization of this principle to describe the behaviour of electrons in asymmetric QHS in tilted magnetic fields is a very delicate question requiring separate grounds. Indeed, the presence of macroscopic-range correlations in the system as well as the presence of a macroscopic ordering of electrons clearly require a new theoretical approach to the system description especially regarding the selection of the boundary conditions. We anyway hope that the number of intriguing effects as well as the number of unresolved problems will stimulate further activity in the field of asymmetric QHS.

In summary, we have directly demonstrated that in asymmetric quantum Hall system under tilted magnetic fields a local photo-response is a complicate non-monotonic function of macroscopic distances to the sample edges. We have also observed the effect of a local photo-response in the dark caused by illumination of the sample arias remote on a macroscopic distance. These macroscopic-range correlations indicate the presence of a macroscopic ordering of electrons in the system.

**Acknowledgements**
.
The MBE samples were kindly provided by B.Ya. Meltser and S.V. Ivanov, Center of Physics of Heterostructures, A.F.Ioffe Institute. Partial support of the Russian President's Foundation (Grant No. SS-5920.2006.2) is acknowledged.

**Figures**

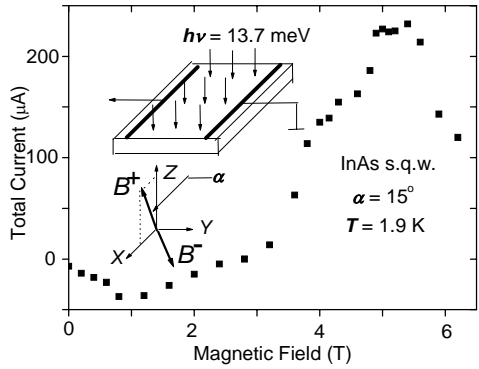

Fig. 1. Total light-induced current as a function of tilted magnetic field at $B = B^+$. The inset shows geometry of the experiment as well as the relative orientation of magnetic field which may be switched from $B = B^+$ to $B = B^-$.

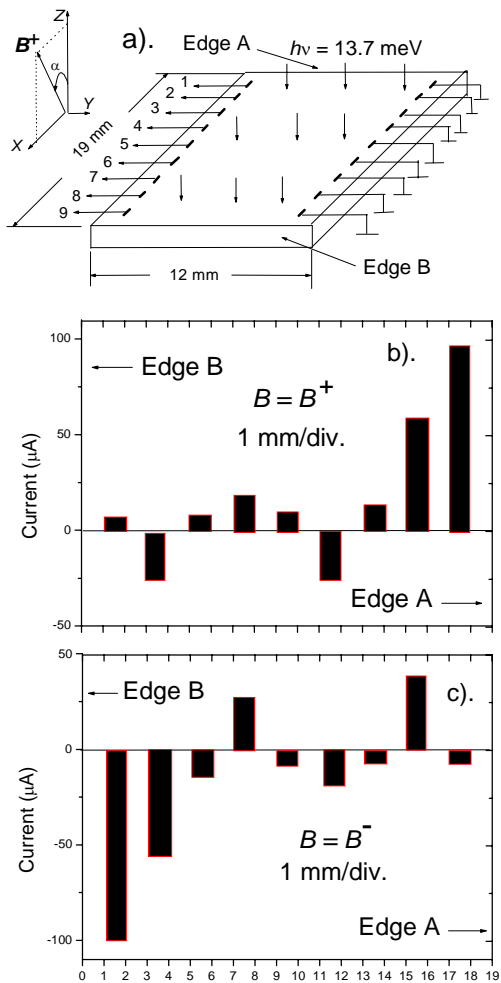

Fig. 2. Distribution of the system's responses on the exposition of the whole sample under CR conditions ($B = 5$ T). Upper panel – geometry of the experiment. Middle panel – distribution of local light-induced currents over the sample length at $B = B^+$. Lower panel – distribution of these currents at $B = B^-$.

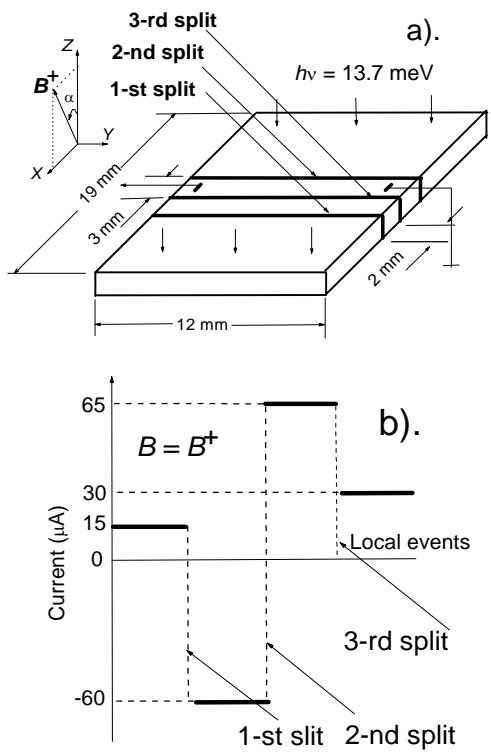

Fig. 3. System's responses on the series of mechanical splits of the sample along $Y$ axis. The whole sample is exposed to light under CR conditions at $B = B^+ = 5$ T. Upper panel – geometry of the experiment. Lower panel – diagram of modifications of the local light-induced current caused by these splits.

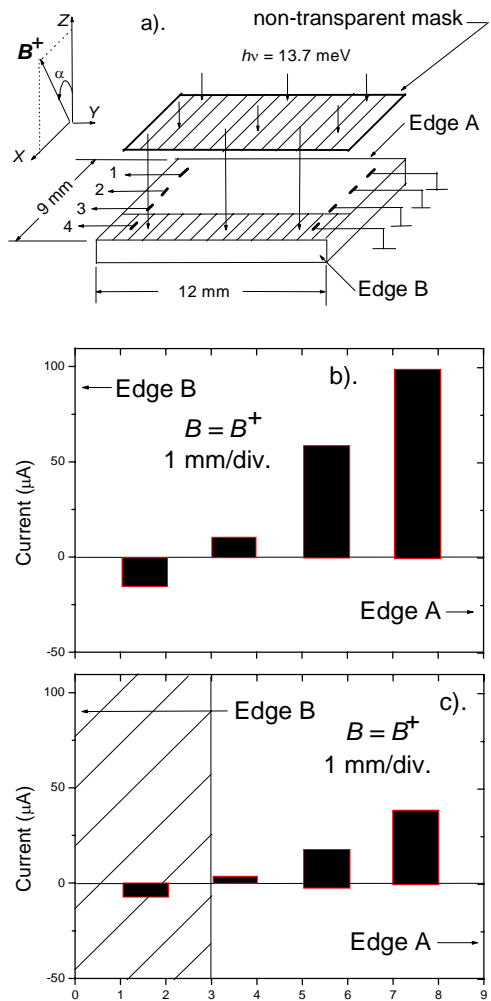

Fig. 4. System's responses on the exposition of different sample regions under CR conditions ($B = B^+ = 5$ T). Upper panel – geometry of the experiment. Middle panel – distribution of local light-induced currents over the sample length in the absence of masking when the *whole* sample is exposed to light. Lower panel – distribution of these currents when *one third* of the sample is exposed to light as it is shown on the upper panel.
In both upper and lower panels, shading denotes those regions which are exposed to light.